\documentstyle[12pt,moriond]{article}
\include{psfig} \psnoisy 

\begin{document}
\heading{THE CONTRIBUTION OF GALAXIES\\ 
         TO THE IR BACKGROUND}

\author{B. Guiderdoni $^{1}$, E. Hivon $^{2}$, \& F.R. Bouchet $^{1}$} 
{$^{1}$ Institut d'Astrophysique de Paris, CNRS, 
98bis Boulevard Arago F--75014 Paris, France.} {$^{2}$ Theoretical 
Astrophysics Center, Juliane Maries Vej 30, DK--2100 Copenhagen, Denmark.}

\begin{moriondabstract}
A new semi--analytic modelling of galaxy evolution in
the IR/submm is hereafter outlined. This type of 
approach successfully reproduces the {\it optical} properties
of galaxies. We illustrate a simple extension to the IR/submm 
wavelength range by taking the case of the SCDM model.
We design a family of evolutionary scenarios with star formation
histories reproducing the evolution of the overall gas and luminosity densities
in the universe. These scenarios are also consistent with IRAS data 
and preliminary ISO counts, and we use 
them to disentangle the ``Cosmic Infared Background'' detected
by Puget {\it et al.} (1996) into discrete sources. We finally
give predictions for the faint galaxy counts and redshift 
distributions in various submm wavelengths, which seem to be
very sensitive to the details of the evolutionary scenarios.
As a consequence, it is easy to anticipate that the current and 
forthcoming instruments, such as SCUBA, FIRST and PLANCK, will 
strongly constrain the evolution of high--$z$ galaxies.
\end{moriondabstract}

\section{Introduction}

The epoch of galaxy formation can be observed by its imprint on
the background radiation which is produced by the accumulation
of the light of extragalactic sources along the line of sight. 
The search for the ``Cosmic Optical Background''
(hereafter COB) currently gives only upper limits. However, the shallowing
of the faint counts obtained in the Hubble Deep Field (HDF,
Williams {\it et al.} 1996) suggests that 
we are now close to
convergence. Thus an estimate of the COB can be obtained by summing up the 
contributions of faint galaxies. At larger wavelengths, the DIRBE 
instrument on COBE has only given
upper limits on the FIR background at 2 -- 300 $\mu$m, while
Puget {\it et al.} (1996) have 
discovered an isotropic component in the COBE/FIRAS residuals
between 200 $\mu$m and 2 mm. We shall assume that this is the long--sought
``Cosmic Infrared Background'' (hereafter CIB). Such a detection 
yields the first ``post--IRAS'' constraint on the evolution of high--redshift 
galaxies in the IR/submm range,
before the era of ISO results. As shown in fig. 1, its level 
comparable to the estimate of the
COB obtained by integrating the contributions of faint counts suggests
that a significant fraction of the energy of young stars is absorbed by
dust and released in the IR/submm.

\begin{figure}[ht]
\hbox
{ \psfig{figure=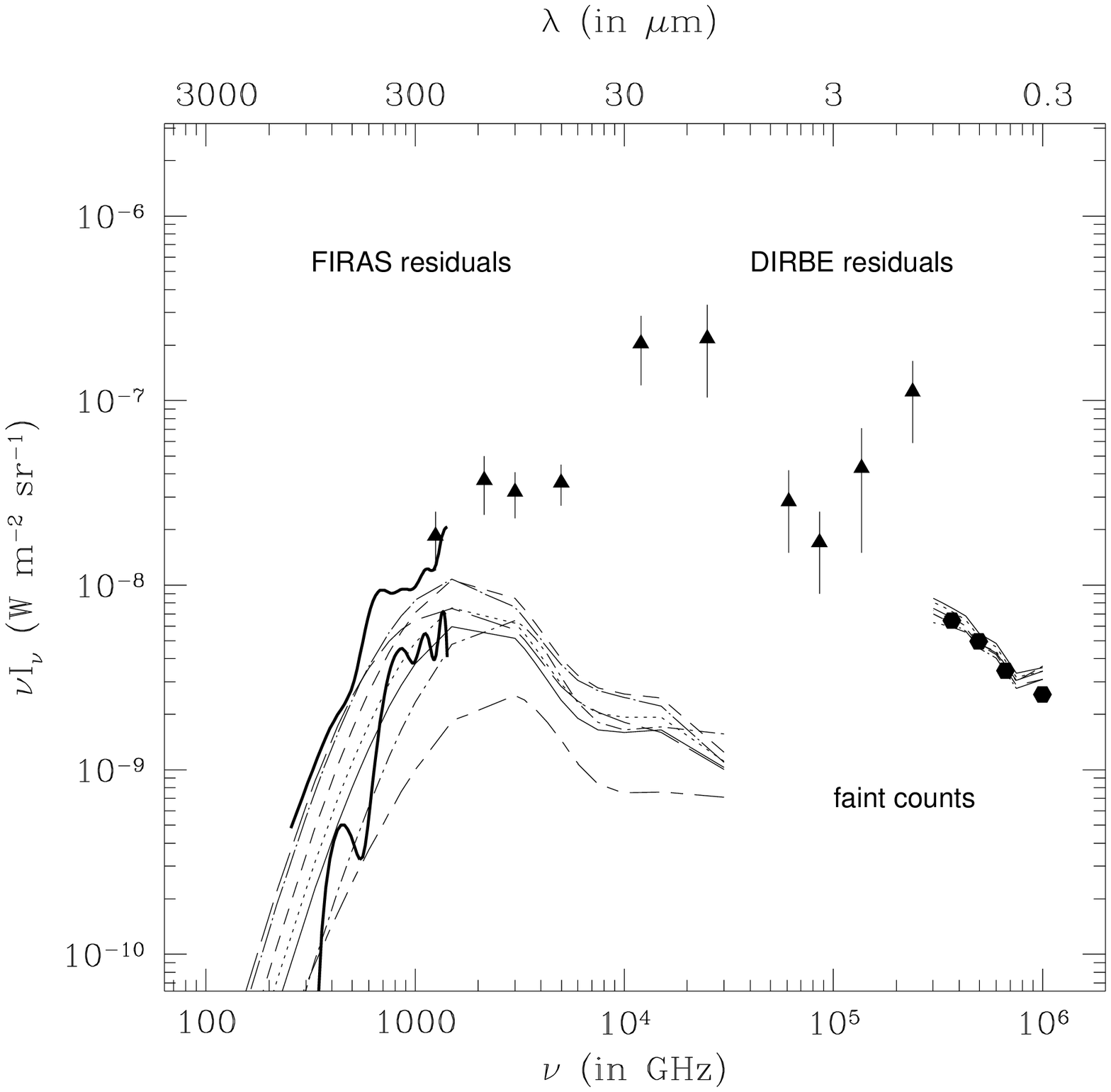,width=0.5\textwidth}
\psfig{figure=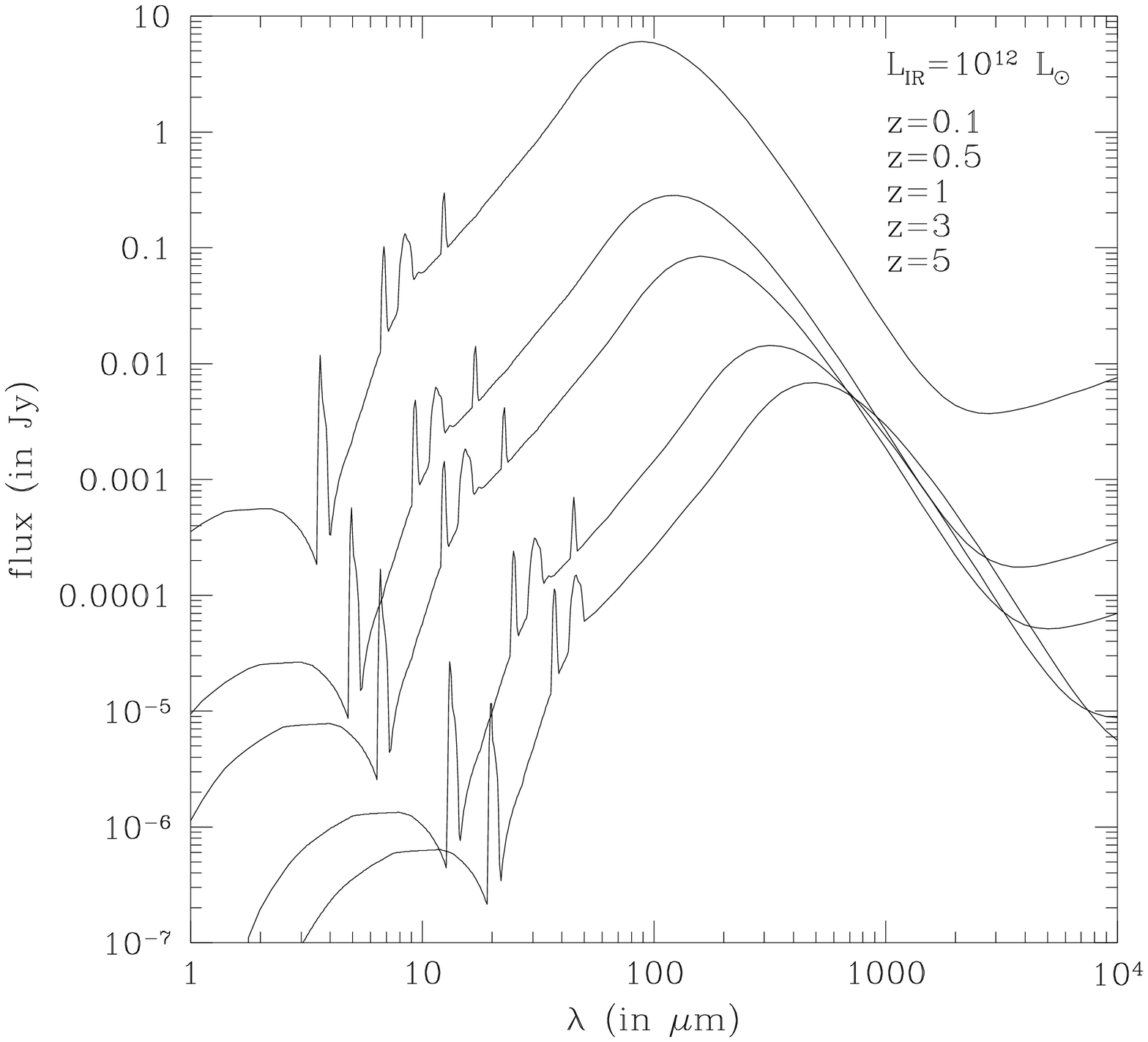,width=0.5\textwidth} }
\caption{\small {\it Left--hand panel:}
Predictions of the diffuse backgrounds in the IR/submm and in 
the optical compared to the current upper limits and detections.
The solid triangles show the level of COBE/DIRBE residuals from Hauser (1995).
The similar shapes of the residuals and dark sky suggest that the
substraction of foregrounds has been incomplete and that the plotted values
are only upper limits.  
The thick solid lines give the CIB detection at $\pm 1\sigma$ 
per point from the re--analysis of the COBE/FIRAS residuals, initiated in
Puget {\it et al.} (1996) and revisited in Guiderdoni {\it et al.} (1997a).
The solid hexagons show the Cosmic Optical Background (COB) obtained by
summing up faint galaxy counts down to the Hubble Deep Field limit. 
The small dashes and long dashes give the prediction for 
no--evolution integrated up to redshift $z_{for}=8$ in a cosmology 
with $h=0.5$ and $\Omega_0=1$.
The other curves are computed for the SCDM model with $h=0.5$, 
$\Omega_0=1$, $\sigma_8=0.67$. The scenarios of tab. 1
are plotted with the following line codes:
Q (dots and small dashes), A (solid line), B (dotted line), C (long
dashes), D (short dashes), and E (dots and long dashes).
Scenarios Q and A are not 
sufficient to fall within the acceptable range for the CIB.
This suggest the existence of an additional population of ULIGs
taken into account in scenarios B, C, D, and E.
{\it Right--hand panel:}
Observer--frame model spectra of a $L_{IR}=10^{12} L_{\odot}$
galaxy at increasing redshifts (from top to bottom).
The reader in invited to note that the apparent flux in 
the submm range is almost insensitive to redshift, because the shift of the
100 $\mu$m bump counterbalances the distance dimming.}
\end{figure}

At the moment, our understanding of galaxy evolution is mainly based on
the recent breakthroughs brought by UV/visible observations
(Lilly {\it et al.} 1995; Steidel {\it et al.} 1996;
Williams {\it et al.} 1996). These results nicely match 
a scenario where star formation in bursts triggered by interaction/merging
consumes the gas content of galaxies as time goes on. But the total amount
of energy released by stars should be estimated by summing
up the UV/visible light of stellar populations directly
escaping from the galaxy, and the part which has been absorbed by dust 
and re--emitted in the IR/submm wavelength range.
The corrections
needed to account for dust extinction are rather uncertain and might induce an
upward revision of the high--redshift star formation rates (SFR) 
deduced from UV/visible observations (Madau {\it et al.} 1996)
by factors of a few. Moreover, a significant fraction of star formation
might be completely hidden in heavily--extinguished galaxies which are
missed by the above--mentionned observations.
While we have learned from IRAS that about
one third of the bolometric luminosity in the local universe is released
in the IR/submm (Soifer and Neugebauer 1991), we know very little
about galaxy evolution in this wavelength range. Faint galaxy counts 
and redshift surveys down
to flux densities $S_\nu \sim 60$ mJy (at 60 $\mu$m) do not
probe deeper than $z\sim 0.2$ (Ashby {\it et al.} 1996;
Clements {\it et al.} 1996a).
These surveys seem to show a strong luminosity and/or
density evolution of IRAS sources, but it is difficult to extrapolate
this trend to higher redshifts on a firm ground. 
It is expected that the ISO satellite
will considerably complete and detail this picture. The ISO--HDF counts at 
15 $\mu$m show clear evidence of evolution (Oliver {\it et al.} 1997).

The so--called {\it semi--analytic} approach, which
has been successfully applied to the prediction of the statistical
properties of galaxies, includes the 
dissipative and non--dissipative processes ruling galaxy formation 
in dark matter haloes, complemented by star formation, stellar
evolution and stellar feedback to the interstellar medium 
(White and Frenk 1991;  Lacey and Silk 
1991; Lacey {\it et al.} 1993, Kauffmann {\it et al.} 1993, 1994; Cole {\it et 
al.} 1994; Baugh {\it et al.}
1996, and other papers in these series). It turns out that, 
in spite of differences in the details of 
the models, these studies lead to conclusions 
in the UV, visible and (stellar) NIR which are remarkably similar. 
But none of these models has been applied so far to the prediction
of the properties of galaxies in the IR/submm range.
This is the main aim of this study which uses a simple version
of the semi--analytic approach to disentangle the CIB into discrete units,
and to give 
predictions of faint counts at wavelengths between 60 $\mu$m and 1 mm.
A detailed version of this work is presented in Guiderdoni
{\it et al.} (1997a,b) and Hivon {\it et al.} 1997.

\section{A brief outline of the semi--analytic approach}

\subsection{Non--dissipative and dissipative collapses}

If we assume that the universe is dominated by non--baryonic dark matter,
the formation and evolution of a galaxy in its dark matter halo can be 
briefly sketched as follows:
the initial perturbation, which is gravitationally dominated by non--baryonic
dark matter, grows and collapses. After the (non--dissipative) collapse, 
and subsequent violent relaxation,
the halo virializes, through the formation of a mean potential well
seen by all particles, which consequently share the same 
velocity distribution. 
The shock--heated baryonic component cools and 
collapses at the centre of the potential well. The baryons 
initially have the small rotation velocity 
of the halo created by tidal interactions with other haloes. Because
of angular momentum conservation, their collapse stops when they reach 
rotational equilibrium.

\subsection{Star formation and stellar feedback}

Stars begin to form in this disk-like baryonic
core and evolve through the main stages of stellar evolution. 
Observations seem to show that the SFR per unit surface density
is proportional to the total gas surface density (neutral plus
molecular) divided by the dynamical time scale of the disk
(Kennicutt 1997). So we shall hereafter
assume that the star formation rate is $SFR(t) = M_{gas}(t)/t_\star$
with $t_\star \equiv \beta t_{dyn}$, the dynamical time scale of the disk.
The free parameter $\beta=100$ accomodates the 
histogramme of ``Roberts times'' observed for a sample of 63 bright disk
galaxies by Kennicutt {\it et al.} (1994). 
The Initial Mass Function is Salpeter's. The spectrophotometric evolution,
gas content and metallicity are consistently followed as in Guiderdoni
and Rocca--Volmerange (1987), with upgraded stellar data.

The feedback can be local or non--local. Local feedback is due to 
supernova explosions and introduces a cut--off in the fraction
of stars that form before the triggering of the galactic wind
$F_\star = (1 + (V_{hot}/V_c)^\alpha)^{-1}$. Cole {\it et al.} (1994)
give the values $V_{hot}=130$ km s$^{-1}$
and $\alpha=5$ from numerical simulations. 
Non--local feedback is due to reheating
(Efstathiou 1992) at $z \geq 2$ in haloes with $V_c \leq V_{equ} \equiv
20$ -- 50 km $^{-1}$ and possibly as high as $V_c \leq (200)^{1/3}
V_{equ}$ in case of adiabatic collapse. We choose to model the two feedbacks by
taking a somewhat arbitrary $(1+z)$ dependence 
$V_{hot}=40(1+z_{coll})$ km s$^{-1}$. 

\subsection{Dust absorption and emission}

Part of the energy released by stars is absorbed by dust 
and re--emitted in the IR and submm ranges. 
The derivation of the IR/submm spectrum is a three--step process:
i) Computation of the optical thickness of the disks.
We assume that the gas is distributed in an exponential disk
with truncation radius $r_g$ (computed from $r_g/r_{25}=1.6$, Bosma 1981).
As in Guiderdoni and Rocca--Volmerange (1987),
the extinction curve depends on the gas metallicity $Z_g(t)$ 
according to power--law 
interpolations based on the Solar Neighbourhood and the Magellanic Clouds.
ii) Computation of the amount of bolometric energy absorbed by dust. 
We assume a simple geometric 
distribution where the gas and the stars which contribute mainly to
dust heating are distributed with equal scale heights in the disks
(the so--called ``slab'' geometry). We average over inclination angle $i$
and crudely take into account the effect of the albedo $\omega_\lambda$ for 
isotropic scattering (Natta and Panagia 1984).
iii) Computation of the spectral energy distribution of dust emission.
The emission spectra are computed as a sum of various components,
according to the method developped by Maffei (1994), from the observational
correlations of the IRAS flux ratios 12$\mu$m/60$\mu$m, 25$\mu$m/60$\mu$m
and 100$\mu$m/60$\mu$m with $L_{IR}$ (Soifer and Neugebauer 1991). 
These correlations are extended to low $L_{IR}$ with the sample of
Rice {\it et al.} (1988). Several components are considered in the model 
spectra (according to D\'esert {\it et al.} 1990):
Polycyclic aromatic hydrocarbons (PAH), very small grains (VSG) and
big grains (BG). Synchrotron radiation is strongly correlated
with the IR luminosity
(see e.g. Helou {\it et al.} 1985). We extrapolate its contribution
from 21 cm down to $\sim $ 1 mm with a single average slope 0.7.

Fig. 1 shows observer--frame {\it model} spectra of a
luminous IR galaxy at various redshifts. It turns out that
there is a wavelength range, between $\sim 600$ $\mu$m and $\sim 4$ mm, where 
the distance effect is counterbalanced by the ``negative k--correction'' 
due to the huge rest--frame emission bump at $\sim 100$ $\mu$m. In 
this range, the apparent flux of galaxies
depends weakly on redshift. The modelling of the observer--frame submm
fluxes, faint galaxy counts and diffuse background
of unresolved galaxies is consequently very sensitive to the early stages
of galaxy evolution.

\section{The history of star formation}

\subsection{Two modes of star formation}

Hereafter, we consider the SCDM model with $H_0=50$ km s$^{-1}$ 
Mpc$^{-1}$, $\Omega_0=1$, $\Lambda=0$, $\sigma_8=0.67$, and
$\Omega_{bar}=0.05$ as an illustrative case.
We need to introduce scenarios of evolution which will
be used to compute the IR/submm properties of galaxies and to predict the
CIB and faint galaxy counts. 
While a complete assessment of the energy budget of 
galaxies would require the monitoring of the multi--wavelength
luminosity functions and 
galaxy counts (delayed to forthcoming studies), we hereafter only wish to 
address the issue of the overall evolution of the comoving
gas (and SFR) density in the universe. 

Fig. 2 shows the predicted gas evolution for scenario Q
with $\beta=100$. It corresponds to the fit of SFR timescales in disks, 
that is, the so--called ``Roberts times'' peaking at 3 Gyr and ranging between 
0.3 and 30 Gyr (Kennicutt {\it et al.} 1994). 
Clearly the gas density does not decline fast enough between
$z=1$ and the local universe because the SFR is not high enough.
It is not surprising to check in fig. 1 that
the predicted background is below the acceptable range for the CIB.

\begin{figure}[hp]
\hbox
{ \psfig{figure=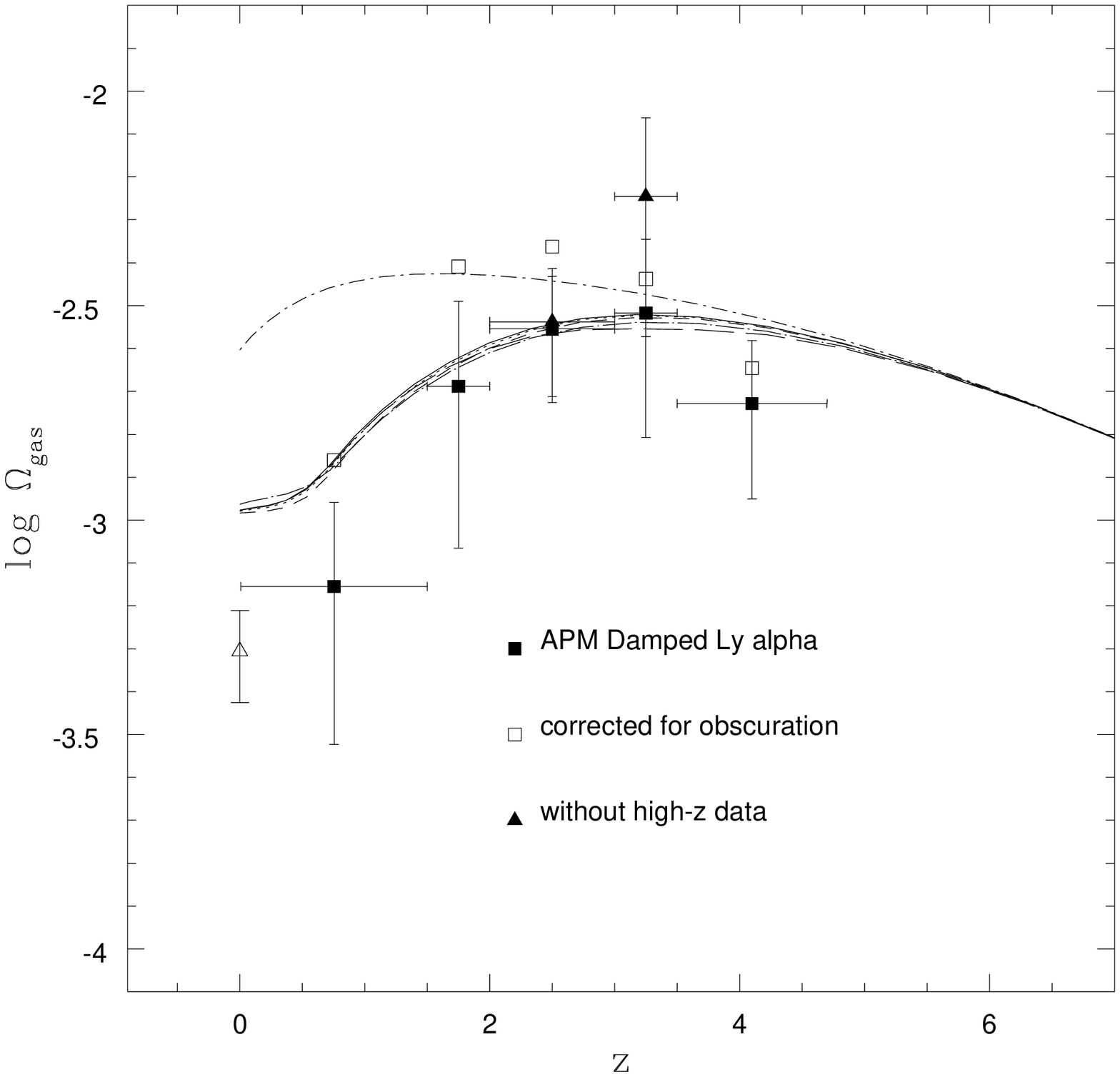,width=0.5\textwidth} 
\psfig{figure=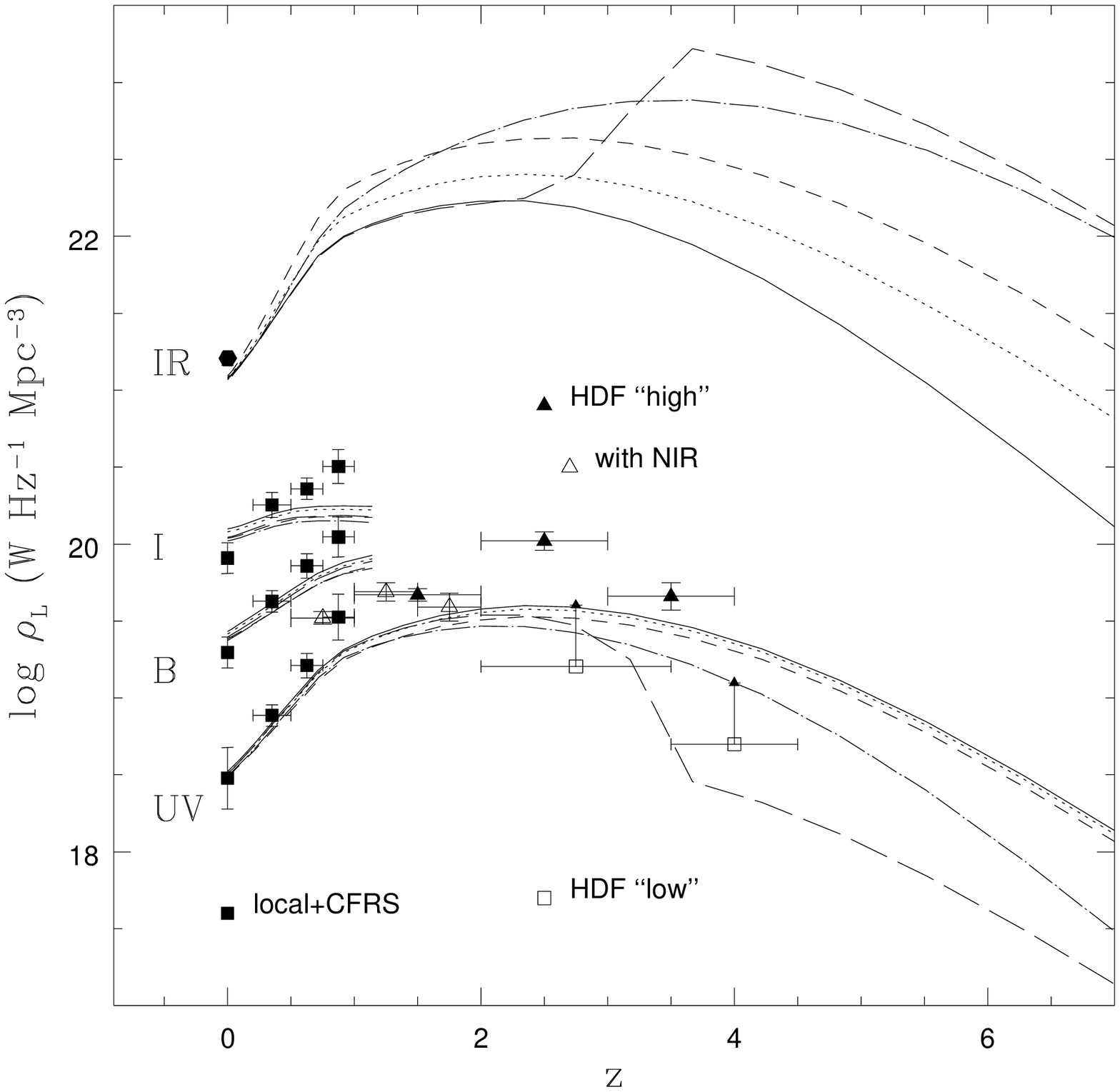,width=0.5\textwidth} }
\caption{\small {\it Left--hand panel:}
Evolution of the cold gas
density parameter in damped Lyman--$\alpha$ systems. Solid triangles: data
without the APM QSO survey. Solid squares: data including the APM
QSO survey. Open squares: tentative correction for selection effects due to
QSO obscuration (Storrie--Lombardi {\it et al.} 1996). Open triangle: 
local estimate from HI surveys (Briggs and Rao 1993). 
Scenario Q (``quiet'' mode, $\beta=100$) is plotted with dots 
and small dashes. Other scenarios (plotted with line codes of fig. 1)
involve a ``burst'' mode
($\beta=10$) fraction increasing with redshift. The various scenarios involving
the ``burst'' mode consume more gas than the ``quiet'' mode.
{\it Right--hand panel:}
Rest--frame comoving luminosity densities. Letters
UV, B, I and IR respectively stand for 2800\AA, 4400
\AA, 10000 \AA\ and 60 $\mu$m. 
The emissivity at 1600 \AA\ is about 30 \% higher than at 2800 \AA.
Solid squares: local and Canada--France Redshift Survey (Lilly {\it et al.} 
1996). Open triangles: NIR data are taken into account to
compute photometric redshifts in the Hubble Deep Field 
(Connolly {\it et al.} 1997). Solid triangles: other estimates of photometric 
redshifts in the HDF (Sawicki {\it et al.} 1997). Open squares: HDF with 
redshifts from Lyman--continuum drop--outs (Madau {\it et al.} 1996).
Solid hexagon: 60 $\mu$m local density corresponding to one third of 
the bolometric light radiated in the IR (Saunders 
{\it et al.} 1990).  
Scenario A (solid line) has no ULIGs. Various
quantities of ULIGs are included in scenarios B (dotted line), C (long
dashes), D (short dashes), and E (dots and long dashes). See tab. 1 for 
details.
The different UV and IR emissions mainly result from different IMF and
extinction, with almost similar SFR histories, and they are not strongly
constrained by the 
current status of (possibly discrepant) UV/visible observations.}
\end{figure}

We now consider a mix of two broad types of populations,
one with a ``quiet'' star formation rate, the other proceeding in bursts
with $\beta=10$. For the ``burst'' mode, we take an
involved mass fraction increasing with collapse redshift 
$f_{burst}(z)=f_{burst}(0)(1+z_{coll})^\gamma$, as suggested by the increasing
fraction of blue objects showing tidal and merger features at larger $z$
(Abraham {\it et al.} 1996). 
Noting that the frequency of galaxy pairs is $\propto (1+z)^\delta$ with
$\delta$ ranging between 2 and 6 at $\pm 1 \sigma$ 
(Zepf and Koo 1989; Burkey {\it et al.} 1994;
Carlberg {\it et al.} 1994), we choose here a high evolution rate $\gamma
=5$. Then $f_{burst}(0)$ is set to 0.04 in order to fit the SFR density at 
low $z$, resulting in an ``all--burst'' behaviour at $z \geq 0.8$.
This will be our scenario A.
As shown in fig. 2, this phenomenological description of the increasing
importance of the bursts reproduces the steep decline of the gas density. The 
origin of this fast evolution has still to be elucidated by
a more exhaustive modelling of all interaction processes in semi--analytic
codes following the merging history trees of haloes and galaxies.

We can now compute the corresponding IR/submm emission
by taking the conservative estimate of the average optical thickness
and ``slab'' geometry as in the ``quiet'' mode. As a result, this population of
``mild starbursts'' and ``luminous UV/IR galaxies'' (LIGs) has IR--to--blue 
luminosity ratios in the range $0.06 \leq L_{IR}/\lambda_B L_B \leq 4$
which is characteristic of blue--band selected samples (Soifer {\it et al.} 
1987), and should be fitted to the
Canada--France Redshift Survey (CFRS,
selected in the observer--frame $I_{AB}$ band,
roughly corresponding to the $B$ band at $z \sim 1$), and to high--$z$
HDF galaxies.
The evolution of the comoving luminosity density in various UV/visible
bands and at 60 $\mu$m is compared to
observational estimates in fig. 2. The local energy budget and its evolution 
from $z=0$ to 1 seem to be fairly reproduced. 
As shown in fig. 1, scenario A reproduces the COB,
and gives a predicted IR/submm background which is clearly 
barely compatible with
the observed CIB whose mean amplitude is twice the prediction, despite our high
choice of $\gamma$. Consequently, we can easily suspect the existence
of a population of galaxies which are more heavily extinguished.

\begin{table}[htbp]
\caption{Scenarios of galaxy evolution}
\begin{center}
\begin{tabular}{@{}lrrr} \hline
Name & $f_{burst}$ & $f_{quiet}$  & \% of ULIGs \\
     & ($\beta=10$)&($\beta=100$) &            \\
Q    & 0                        & 1                      & 0 \% \\
A    & $0.04(1+z_{coll})^5$   & $1-f_{burst}$           & 0 \% \\
B    & $0.04(1+z_{coll})^5$   & $1-f_{burst}$           & 5 \% at all $z$ \\
C    & $0.04(1+z_{coll})^5$   & $1-f_{burst}$  & 90 \% for $z_{coll} > 3.5$ \\
D    & $0.04(1+z_{coll})^5$   & $1-f_{burst}$           & 15 \% at all $z$ \\
E    & $0.04(1+z_{coll})^5$   & $1-f_{burst}$  & $1- \exp -0.02(1+z_{coll})^2$ \\
\end{tabular}
\label{ta:mod}
\end{center}
\end{table}

\subsection{A heavily--extinguished component}

The above-mentionned CIB computed with scenario A
seems to be sort of a
conservative estimate of the minimum IR/submm background
due to typical CFRS and HDF galaxies.
We now wish to assess how much star formation might be hidden by dust shrouds
and introduce an additional population of heavily--extinguished bursts, 
which are similar to ``ultra-luminous IR galaxies'', or ULIGs (Sanders
and Mirabel 1996; Clements {\it et al.} 1996a,b). We maximize their IR
luminosity by assuming that all the energy available from stellar
nucleosynthesis ($0.007xMc^2$) is radiated in a heavily--extinguished medium.
We take $<x>=0.40$ for stars with masses larger
than $\sim 5 M_\odot$.  We
distribute this population of ULIGs in two ways: i) A constant mass 
fraction of 5 \% (scenario B) or 15 \% (scenario D) at all $z$, 
mimicking a scenario of continous bulge formation as the
end--product of interaction and merging. 
ii) 90 \% of all galaxies forming at high $z_{coll}
\geq 3.5$ undergo a heavily--extinguished burst, mimicking a strong episode of
bulge formation. These scenarios are now able
to fit both the COB and CIB as shown in fig. 1. Scenarios B and D seem more
appropriate to reproducing the CIB at 300 $\mu$m while
scenario C, with high--$z$ ULIGs, has a stronger contribution at
larger wavelengths. Of course, these last three cases are only illustrative,
and a combination of these solutions would also fit the
CIB. For instance, we introduce
an ad hoc scenario E with a fraction of ULIGs increasing as
$1- \exp -0.02(1+z_{coll})^2$. Such a dependence can be obtained if the 
fraction
of ULIGs depends on the mean surface density and optical thickness
of disks which roughly scale as $(1+z_{coll})^2$ in our modelling.
Scenario E nicely fits the COB and CIB.

\begin{figure}[htpb]
\vbox {
\hbox {
\psfig{figure=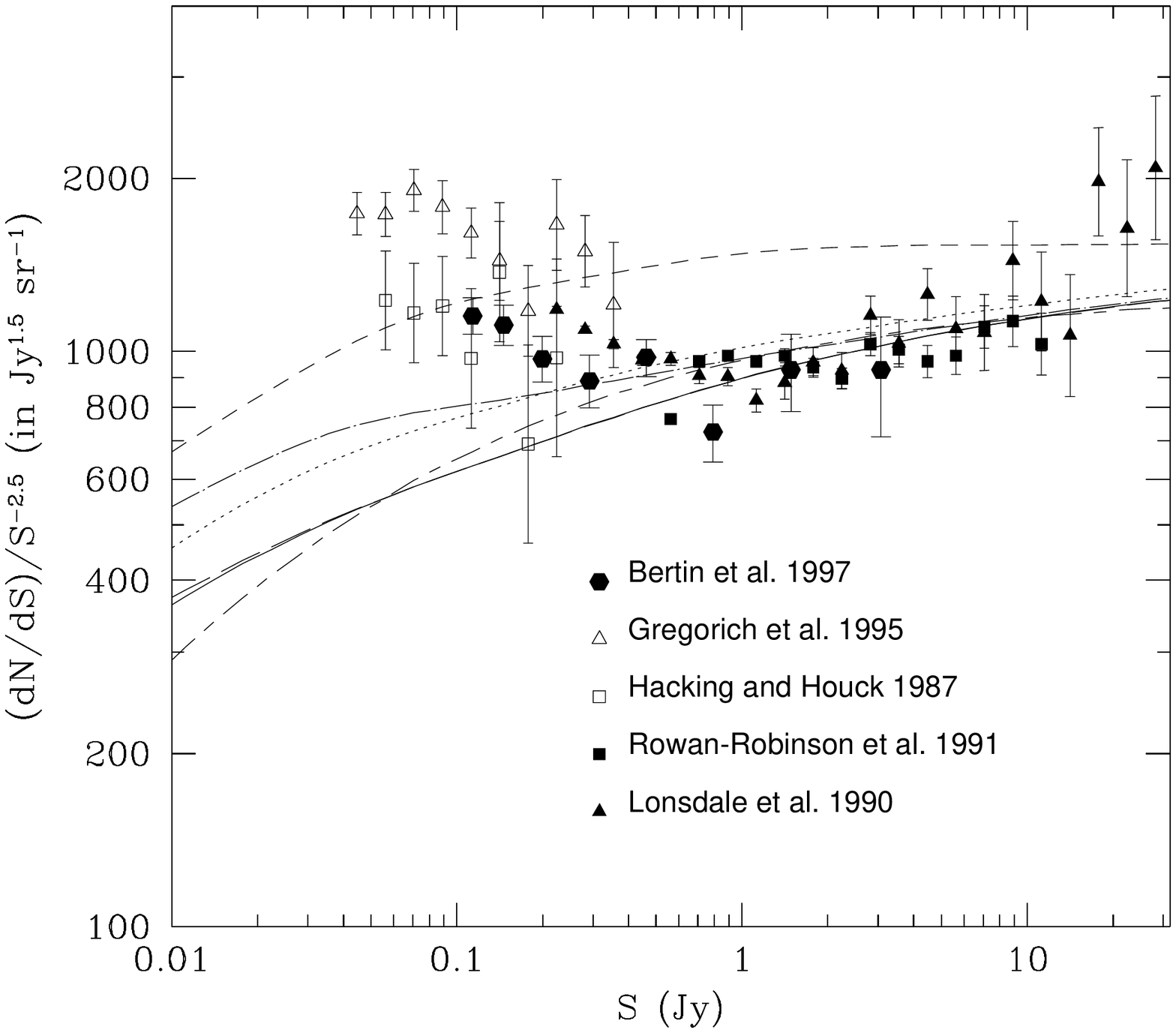,width=0.5\textwidth}
\psfig{figure=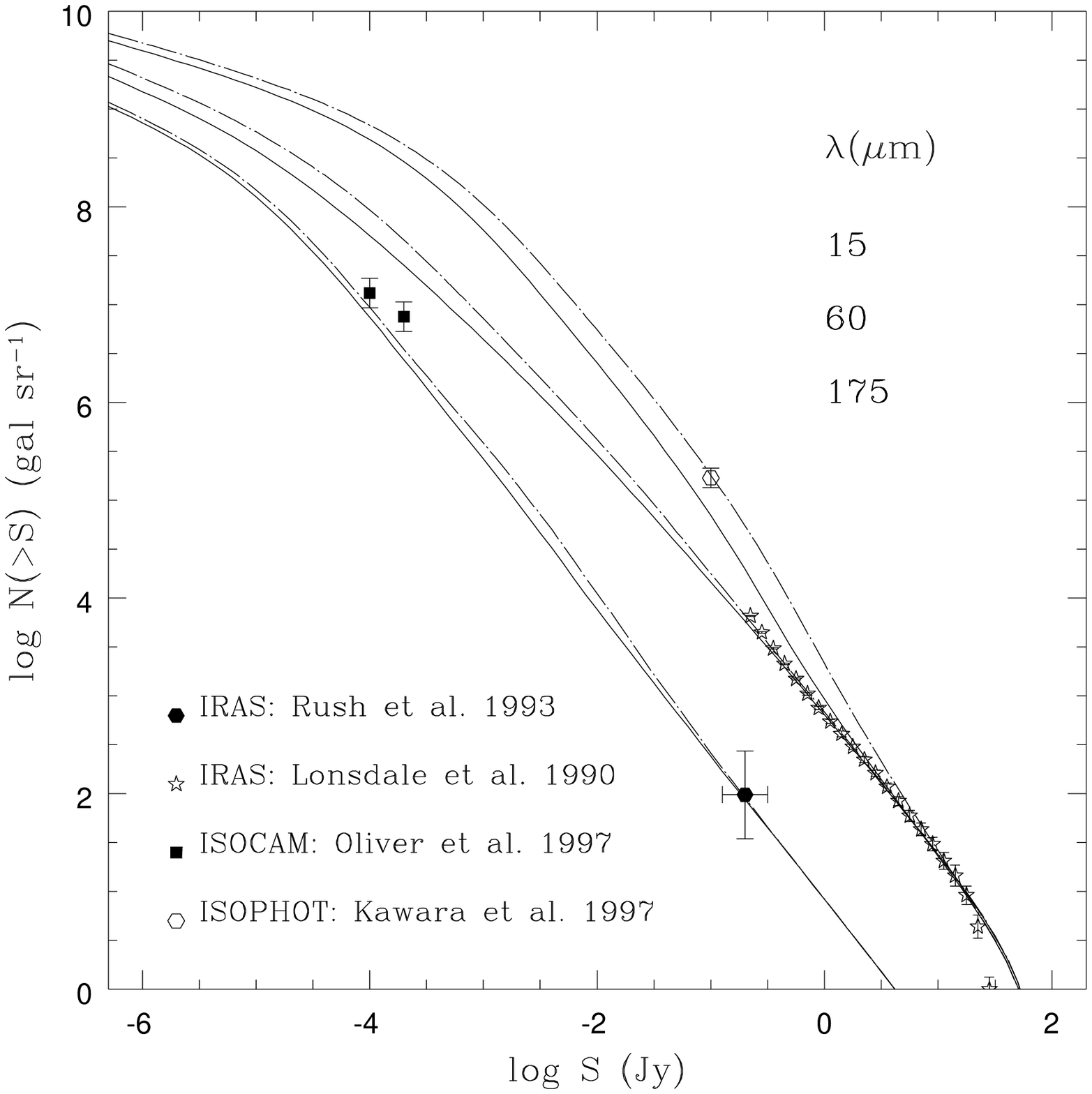,width=0.5\textwidth} }
\hbox {
\psfig{figure=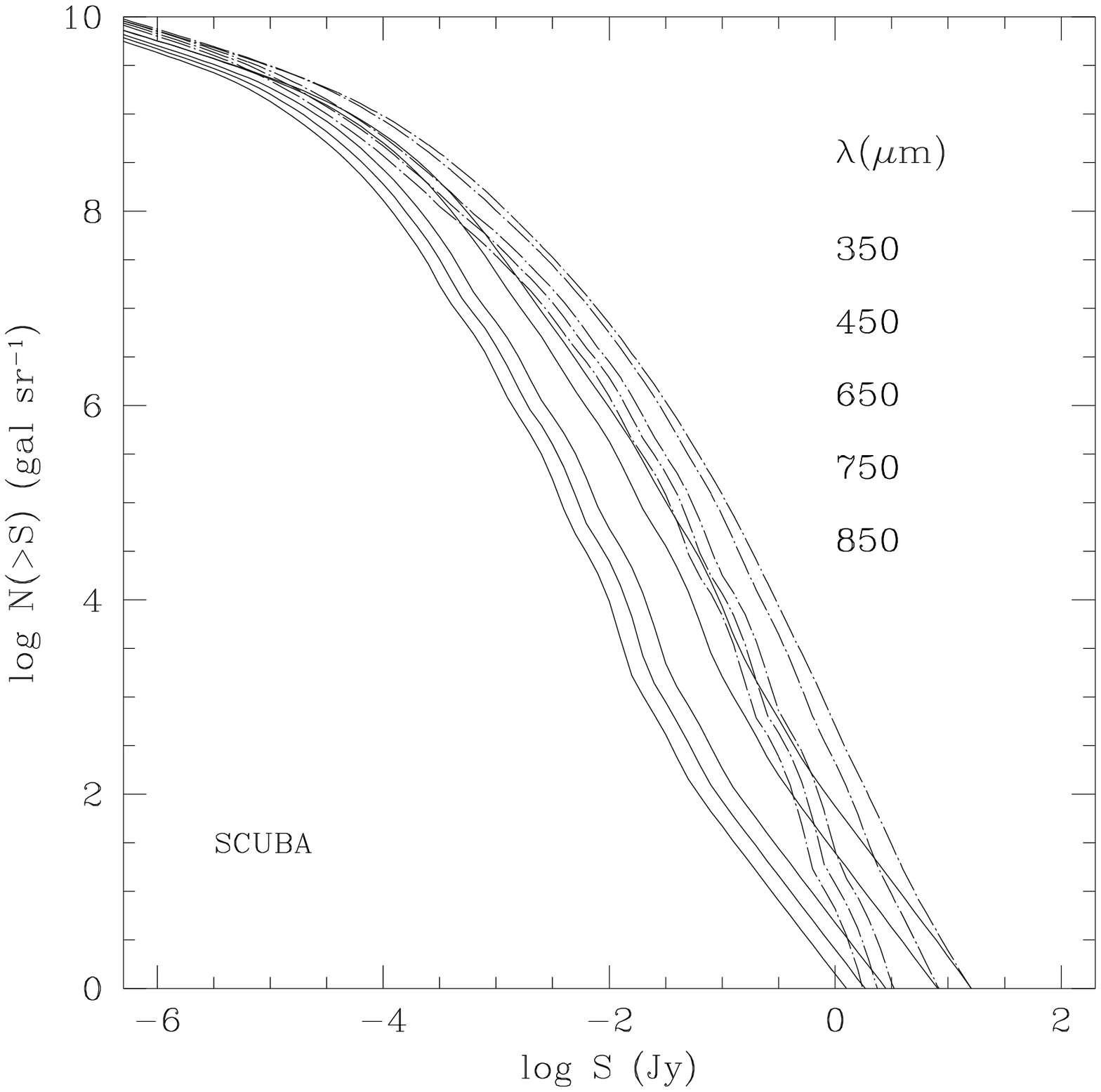,width=0.5\textwidth}
\psfig{figure=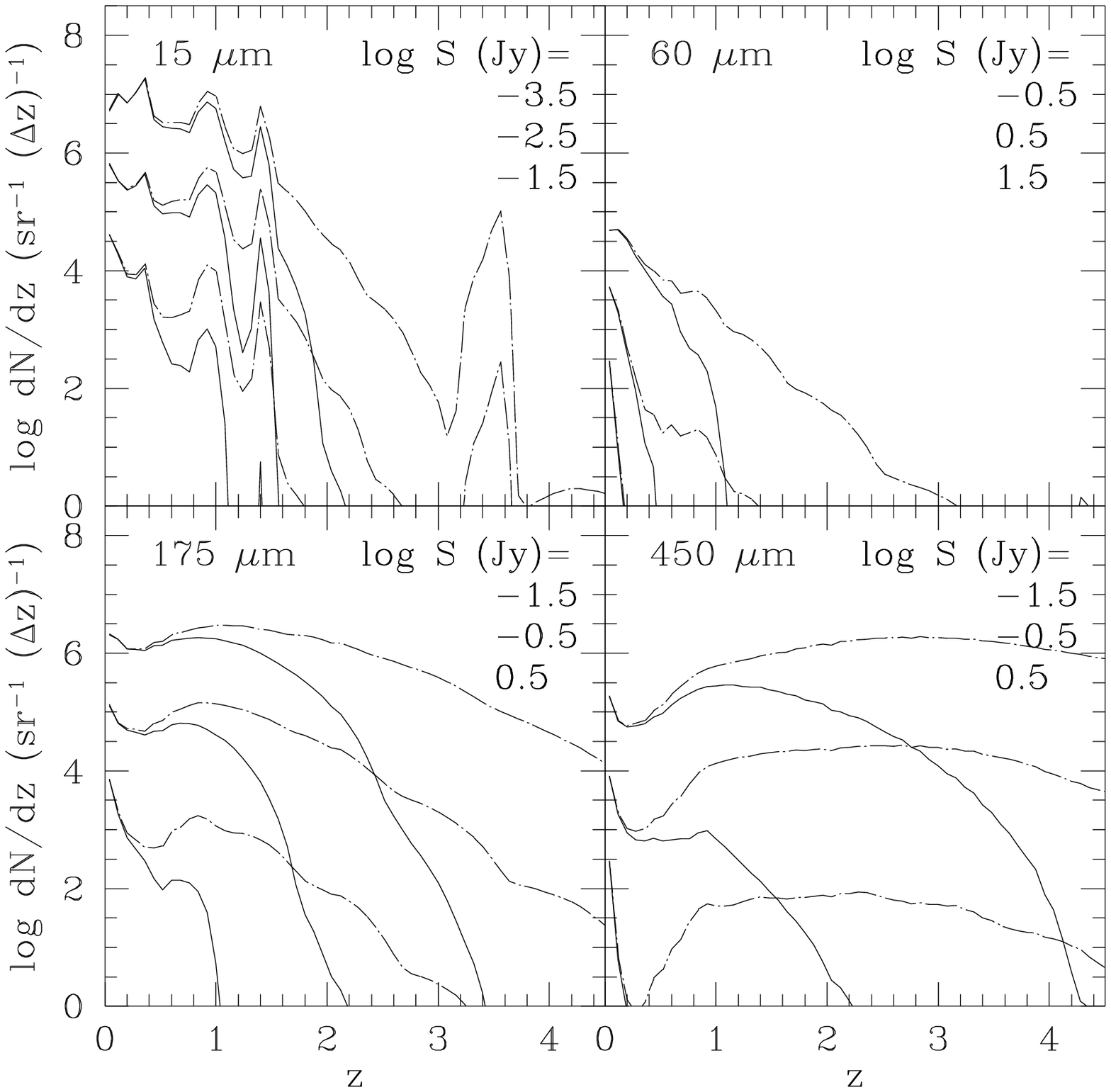,width=0.5\textwidth} }
}
\caption{\small 
{\it Left--hand, top panel:}
Predictions for differential galaxy counts (normalised to 
Euclidean counts) at 60 $\mu$m. Data is shown for
IRAS counts. Open stars: Faint Source Survey
(Lonsdale {\it et al.} 1990). Open squares: QMW survey 
(Rowan--Robinson {\it et al.} 1991).
Solid squares: North Ecliptic Pole Region 
(Hacking and Houck 1987). These latter counts might be affected by the 
presence of a supercluster. Solid hexagons: Very Faint Source Survey
(Bertin {\it et al.} 1997).
The counts by Gregorich {\it et al.}
(1995), which are plotted with open triangles, are contaminated by cirrus.
The no--evolution model (for a cosmology with $h=0.5$, $\Omega_0=1$)
is shown with the short and long dashes. Scenarios A, B, C, D, E
are plotted with line codes as in fig. 1. 
{\it Right--hand, top panel:} Predictions for faint counts at 15 $\mu$m, 
60 $\mu$m and 175 $\mu$m (from bottom to top)
for scenarios A and E. Note that the 15 $\mu$m fluxes only include dust 
emission, and are lower limits for $z \geq 1$. Open stars: 
Faint Source Survey (Lonsdale 
{\it et al.} 1990). Solid hexagon: Rush {\it et al.} 1993. Solid squares:
ISO--HDF (Oliver {\it et al.} 1997) with ISOCAM at 15 $\mu$m. Open
hexagon: preliminary counts by Kawara {\it et al.} (1997) with ISOPHOT
at 175 $\mu$m.
{\it Left--hand, bottom panel:} Predictions at SCUBA wavelengths for
scenarios A and E. In contrast with counts at wavelengths
smaller than 100 $\mu$m, the submm counts are very sensitive to the
details of the high--$z$ evolution, because of the shift of
the 100 $\mu$m bump into the observing bands.
{\it Right--hand, bottom panel:} Predictions of redshift distributions
in various bands for scenarios A and E. The central values of the 
$\Delta \log S_\nu =1$ flux bins (in Jy) are indicated from top to bottom.}
\end{figure}

It is clear from fig. 2 that none of the optical data reflects the 
large differences between these scenarios,
although the fraction of light in the IR varies widely at high $z$.
In scenario A, the IR/UV ratio decreases with increasing $z$ because of
the decreasing metallicity of galaxies. In scenario B and D, this effect in 
cancelled because the ULIG bursts are assumed to be optically--thick,
and the IR/UV ratio at $z=4$ is similar to that at $z=0$.
In scenario C and E, the IR/UV ratio strongly increases with $z$ and is 
$\sim 100$ times higher than for model B at $z=4$. One should note 
that at this redshift, scenario C and E are roughly consistent 
with the lower limits of the UV--luminosity density derived from 
Lyman--continuum 
drop--outs, but with ten times as much SFR as directly derived if
extinction is not taken into account. In these scenarios, galaxy formation 
at high $z$ is an almost completely--obscured process.

\section{Predictions of FIR and submm counts}

Fig. 3 gives the predictions for the IRAS 60 $\mu$m differential counts
normalised to the Euclidean slope. 
All the scenarios involving an increasing fraction
of the ``burst'' mode predict more galaxies than the no--evolution curve.
Scenario D with 15 \% of ULIGs is rejected by the data. All the other
scenarios have a moderate local fraction of ULIGs and are in agreement with 
the faint counts. Scenario E with an increasing fraction of ULIGs gives
an almost flat curve which is reminiscent of the observational trend.
Fig. 3 also gives predicted faint counts at 15 $\mu$m (compared with
the ISO--HDF, Oliver {\it et al.} 1997) and at 175 $\mu$m (compared with
preliminary ISOPHOT counts, Kawara {\it et al.} 1997).
Guiderdoni {\it et al.} (1997a) emphasized the relative degeneracy of the
predictions at wavelengths shorter than 100 $\mu$m, including the IRAS 60 
$\mu$m and ISOCAM 15 $\mu$m counts, and the strong 
sensitivity of the submm counts to the details of galaxy evolution.
While the IRAS data do not yield tight constraints
on the evolution of galaxies at $z \geq 0.2$,
the on--going deep surveys with ISO and the forthcoming ones
with SCUBA in the
atmospheric windows at 450 and 850 $\mu$m (for which the array is matched 
to the instrument optics) will quickly help discriminate between the various
scenarios, in the expectation of PLANCK and FIRST.
The 10 mJy level which will
be reachable by the forthcoming submm observations with SCUBA and FIRST
will allow the surveys
to begin ``breaking'' the CIB into discrete units. As shown in fig. 3,
a large fraction of these objects are at redshifts beyond 1.  

\section{Conclusions}

Predictions of IR/submm galaxy counts, redshift distributions,
and diffuse background of unresolved galaxies can be
obtained from a semi--analytic model of galaxy formation and
evolution which takes into account the main
physical processes from the collapse of the density fluctuations to
the absorption of UV and visible star light by dust and the re--emission at
larger wavelengths. This
new model is consistent with the dynamical process of 
continuous galaxy formation predicted by the hierarchical growth of structures
in a SCDM universe, and represents a
significant progress with respect to previous phenomenological models
designed to make predictions in the IR/submm.
 
While the cosmological context and the rules of dissipative collapse 
are fixed, there is some
freedom in the choice of the evolutionary scenarios, through a series
of free parameters describing star formation ($\beta$) and feedback 
($\alpha$ and $V_{hot}$). In this first study, we chose to accomodate
a selected set of data: the comoving SFR, gas and UV/visible luminosity
densities. Since the ``quiet'' mode of star formation ($\beta=100$) 
is unable to 
reproduce the data, we have to introduce a mass fraction involved in the 
``burst'' mode ($\beta=10$) increasing with redshift as $(1+z_{coll})^5$ and 
likely associated to the increasing rate of interaction and merging.
This phenomenological ansatz allows the scenarios to reproduce the evolution
of the overall properties of galaxies between $z=1$ and 0, and has yet to
be explained.  

Then we compute the IR properties of these galaxies. In order to reproduce
both the COB and CIB, it is necessary to introduce an additional population
of heavily--extinguished sources. We design several scenarios
consistent with the CIB and our selected set of data. The IRAS
60 $\mu$m counts are only sensitive to low--$z$ galaxies and weakly
constrain these scenarios, though they seem to limit the fraction of 
local ULIGs. On the contrary, the ISOCAM counts at 15 $\mu$m,
and especially the ISOPHOT and other submm counts are sensitive
to the details of the evolutionary scenarios. The on--going
observations with ISO should already 
discriminate between the range of possibilities which
are still consistent with the IRAS data. Future
observations with SCUBA and FIRST at the 10 mJy level
will help put strong constraints
on the evolution at $z \geq 1$, and begin to ``break'' the CIB into
discrete sources.

\vfill
\end{document}